\documentclass[preprint,showpacs,preprintnumbers,amsmath,amssymb]{revtex4}

\usepackage{graphicx}% Include figure files
\usepackage{dcolumn}% Align table columns on decimal point
\usepackage{bm}% bold math

\graphicspath{{Fig/}}

\begin{document}

\preprint{Manuscript}

\title{The energetic and structural properties of bcc NiCu, FeCu alloys: a first-principles study

%(Monolayer lattice modulation in Quantum Well:  Crystal
%heterogeneity and New interface)
}
\author{Yao-Ping Xie}
\author{Shi-Jin Zhao}
 \affiliation{Institute of Materials Science, School of Materials Science and Engineering, Shanghai University, Shanghai, 200444, China}
\date{\today}% It is always

\begin{abstract}

Using special quasirandom structures (SQS's), we perform
first-principles calculations studying the metastable bcc NiCu and
FeCu alloys which occur in Fe-Cu-Ni alloy steels as precipitated
second phase. The mixing enthalpies, density of state, and
equilibrium lattice parameters of these alloys are reported. The
results show that quasi-chemical approach and vegard rule can well
predict the energetic and structural properties of FeCu alloys but
fail to yield that of NiCu. The reason rests with the difference of
bond energy variation with composition between NiCu and FeCu alloys
induced by competition between ferromagnetic and paramagnetic state.
Furthermore, the calculated results show that the energetic and
structural properties of these alloys can well explain the local
composition of the corresponding precipitates in ferrite steels.

\end{abstract}

% PACS, the Physics and Astronomy
                             % Classification Scheme.
%\keywords{metastable phases,precipitates, first-principles
%calculation, special quasirandom structures(SQS's), magnetism}}
%Use showkeys class option if keyword
                              %display desired

\maketitle

\section{\label{sec:level1}  Introduction}
%\label{}
The transitional metals and their alloys related to the magnetism
attract much scientific interests. Enormous experimental and
theoretical investigations were dedicated to deeper understanding of
the nature of their properties which became complicated due to
magnetism. Examples of these systems can be provided by artificial
crystalline structures fabricated by film growth techniques and
precipitated second phase in matrix. Specially, many important
properties of materials, such as their mechanical strength,
toughness, creep, corrosion resistance, and magnetic properties are
essentially controlled by precipitated particles of a second phase.
Therefore, understanding of these alloys is also desirable from the
views of applications.

The Cu-rich precipitates, which are commonly found in alloy steels,
have good strengthening effect on steels. The high-strength
low-alloy steels strengthened by copper rich precipitates also
retain the impact toughness, corrosion resistance, and welding
properties\cite{NUsteel1,NUsteel2,NUsteel3,NUsteel4,NUsteel5,NUsteel6}.
However, it is also confirmed that the presence of Cu-rich
precipitates in reactor pressure vessel(RPV) steels is the origin of
embrittlement \cite{cu-brittle}, which limit the reactor operating
life. The embrittlement effect can be enhanced by the content of Cu
or Ni of alloyed steels containing of both Cu and
Ni\cite{cu-nin-1,cu-nin-5,cu-ni0new,cu-ni0,cu-ni2}. Thus, the
mechanism of strengthening and embrittlement of Cu precipitates
attract much
interest\cite{ther1,ther2,ther3,ther4,ther5,ther6,ther7,ther8,miller1,miller2}.
The structure and composition of precipitated phase also became
important, and it had been studied by many experiments, such as atom
probe field ion microscopy(APFIM) \cite{a3,a4,a5}, small-angle
neutron scattering(SANS) \cite{a7}, high-resolution and conventional
electron microscopy(HREM and CTEM) \cite{a8, a9}, etc. It was
observed that small Cu-rich precipitates with diameters less than
about 5 nm have a meta-stable body-centered cubic (bcc) structure
and are coherent with the $\alpha$-Fe matrix \cite{a7,a8,a9} in the
initial stage of segregation.  In addition, it was also found that
Ni also occurs in the Cu-rich precipitates beside Cu and Fe. For
HSLA steels, Ni was observed at the coherent matrix/Cu-precipitate
heterophase interfaces \cite{a11,a11-1,FeCu1,FeCu2}. For RPV steels,
it was also observed Ni appeared in the Cu-rich precipitates at very
initial stage and was rejected from the core with the precipitates
growth\cite{new,add5,miller1,miller2}.

The bcc FeCu alloy has been studied intensively by theoretical
investigations for well understanding of precipitated phase.
First-principles calculations based on the cluster expansion
framework gave the composition range of mechanical stability of FeCu
random alloy\cite{a12}. A thermodynamic equilibrium analysis was
used to investigate the composition dependence of Gibbs energy of
FeCu alloy as well as the interfacial energy between precipitates
and matrix\cite{FeCu1}. In the meanwhile, the vibrational energies
of dilute FeCu alloys were investigated by first-principles
calculations in combination with thermodynamical modeling, which
show that the vibrational energies can stabilize the alloys
\cite{FeCu2}. The dependence of magnetism on the structural
characteristics of Cu nucleation and the electronic structure for
bcc Fe$_x$Cu$_{1-x}$($x> 0.75$) systems were also well
investigated\cite{FeCu3}. However few systematic investigation of
bcc NiCu alloys was reported. The bcc NiCu, like bcc FeCu, is not
only a typical system related to magnetism in solids, but also
became interesting because of its important contribution to
strengthening and embrittlement of the ferrite steels. Therefore,
systematic investigations of their properties at electronic level
have both foundational and engineering significance. In this paper,
the composition-dependent bcc Ni$_x$Cu$_{1-x}$ and Fe$_x$Cu$_{1-x}$
random alloys are investigated by employing special quasi-random
structures (SQS's) in the frame of first-principles calculations.

\section{\label{sec:level1}Modeling and Theoretical Methods}

The SQS method was proposed by Zunger and Wei et al to overcome the
limitations of mean-field theories \cite{a13,a14}, without the
prohibitive computational cost associated with directly constructing
large supper cells with random occupy of atoms. The SQS method was
extensively used to study the properties of semiconductor alloys
which are all fcc-based system. Recently, this method was also used
into both fcc and bcc transition metal systems and was proved to be
useful \cite{a15,a16}. For a binary substitutional alloy, many
properties are dependent on its configuration. A binary AB
substitutional alloy with a lattice of $N$ sites has $2^N$ possible
atomic arrangements, denoted as configurations $\sigma$. The
measurable property $<E>$ represents an ensemble average over all
$2^N$ configuration $\sigma$, $<E> = \sum \rho (\sigma) E(\sigma)$.
The structure $\sigma$ can be discretized into its component figure
$f$, which is characterized by a set of correlation functions
$\overline{\Pi}_{k,m}$. Therefore, $<E>$ can be rewrite as $<E> =
\sum \overline{\Pi}_{k,m} \varepsilon(f)$. The SQS's are
 special designed $N$ atoms periodic structures whose distinct
correlation functions $\overline{\Pi}_{k,m}$ best match ensemble
averaged $ \langle \overline{\Pi}_{k,m} \rangle $ of random alloys.

The choice of SQS's are critical for the calculations\cite{new-17}.
In this paper, SQS's containing 32 atoms are constructed for bcc
alloys, which are shown in Fig. 1. The vectors of lattice are
$\overrightarrow{a_1} =( 1.0,-2.0, 0.0)a_0, \overrightarrow{a_2}=(
0.0,-4.0, 2.0)a_0, \overrightarrow{a_3}=(-2.0, 0.0,-2.0)a_0$,
respectively, and $a_0$ is the lattice parameter of the bcc unit
cell. The occupations of these sites for the SQS's at $x$ = 0.25 and
$x$ = 0.5 are given in Table 1, and their structural correlations
function $\overline{\Pi}_{k,m}$ compared with ideal random alloy
correlation functions are given in Table 2. As can be seen, the
quality of the SQS's used in these calculation is reasonably good.

First-principles calculations are performed using the density
functional theory\cite{v43,v43-1,v44} as implemented in the Vienna
ab initio simulation package(VASP)\cite{v49}. The generalized
gradient approximation(GGA)\cite{v47,v48} is used for the exchange
correlation functional. The interaction between core and valence
electrons is described with the projector augmented wave (PAW)
potential \cite{v50,v51}. The equilibrium structures were
determined, up to a precision of 10$^{-4}$ eV in total-energy
difference and with a criterion that required the force on each atom
to be less than 0.01 eV/{\AA} in atomic forces. The convergence
tests with k-point and energy cutoff are presented in table 3.  The
k-point meshes $6 \times 2 \times 4$ for the Brillouin zone
integration are used for SQS's, which is equivalent k-point sampling
with Monkhorst-Pack mesh $11 \times 11 \times 11$ for bcc primitive
cell. An energy cutoff of 280 eV is applied in all cases.

\section{\label{sec:level1}   RESULTS AND DISCUSSIONS }

Firstly, basic properties of bcc Fe, Ni, Cu are presented. The
lattice constant and the magnetic moment for bcc Fe at the
equilibrium volume are 2.83 {\AA} and 2.16 $\mu _B$/atom,
respectively. The lattice constant and bulk modulus for bcc Cu at
the equilibrium volume are 2.89 {\AA} and 130 GPa. These results are
in agreement with earlier studies\cite{a12,feni}. The magnetic
moment for bcc Ni at the equilibrium structure with lattice of 2.81
{\AA} is 0.55 $\mu _B$/atom. The calculated result is well
consistent with previous works using GGA, which predicted bcc Ni is
ferromagnetic\cite{ni-gga1,ni-gga2}. Previous theoretical works
using local density approximation (LDA) predicted bcc Ni is
paramagnetic\cite{feni}. However, it is confirmed that bcc Ni do
possesses a magnetic moment of 0.52 $\mu _B$/atom by experiments
more recently\cite{ni-ex}. In addition, it was proved that the GGA
calculations can describe satisfactorily lattice and elastic
constants of Ni in their observed structures\cite{ni-gga2}.

The mixing enthalpies($\bigtriangleup H$) of bcc NiCu and FeCu
random alloys obtained by first-principles SQS method are shown in
Fig .2. The mixing enthalpy is heat quantity absorbed(or evolved)
during mixing two elements to make homogeneous solid solution, which
  is defined as:
  \begin{equation}     \triangle H =  E_{AB} - x E_A - (1-x) E_B
   \end{equation}
where $E_A$, $E_B$, and $E_{AB}$ are the total energy of A, B, and
AB alloy, and $x$ is the composition of alloys. In the frame of
quasi-chemical approach\cite{book}, the mixing enthalpy is only
determined by the composition-independent bond energy between
adjacent atoms. Hence, mixing enthalpy can be written as:
\begin{equation}
\triangle H= \Omega x(1-x),
\end{equation}
where $\Omega = N_a z
(\varepsilon_{A-B}-\frac{1}{2}(\varepsilon_{A-A}+\varepsilon_{B-B}))$
 is interaction parameter, N$_a$ is Avogadro's number,
 $\varepsilon$ is bond energy, and
$z$ is the number of bonds per atom. Thus, the
composition-independent bond energy between atoms would result a
parabolic variation for mixing enthalpies with compositions.  It can
been seen from Fig. 2, the variation of mixing enthalpies of FeCu
with composition is parabolic, while that of mixing enthalpies of
NiCu exhibit a strong asymmetry and change sign as function of
concentration. These effect were also found for random fcc NiCu
alloys in previous calculations\cite{feni2}.

To study the effect of degree of disorder on the mixing enthalpies
of NiCu, we calculate the ordered structures comparing disordered
structures. Three 4-atoms-super-cells are used to simulate order
alloys with composition $x$ =0.25, 0.50, 0.75. The mixing enthalpies
of these order alloys are represented by solid squares in Fig. 2(a).
It shows the mixing enthalpies of this ordered structures of NiCu
are all lower than those of disordered alloy with the same extent,
and the trend of mixing enthalpies variations with compositions are
the same as that of disorder alloys. The
 mixing enthalpies of order alloy also exhibit
 a strong asymmetry. Therefore, the degree of disorder is
  not the reason of the asymmetry of mixing enthalpies for NiCu alloys.

We turn to investigate bonding property of NiCu and FeCu from
electronic structures for understanding strong asymmetry variation
of mixing enthalpies with composition. The bond energy is critical
factor that determines the mixing enthalpies. The variation of the
bond energy can be reveal from comparison of the density of
states(DOS) of Fe$_x$Cu$_{1-x}$ and Ni$_x$Cu$_{1-x}$. Since the bond
energies of transitional metal are determined by the coupling of
d-band, the d-bands of Cu, Fe, Ni in Fe$_x$Cu$_{1-x}$ and
Ni$_x$Cu$_{1-x}$ with different compositions are plotted in Fig 3.
One can easily understand the difference of mixing enthalpies
magnitude between FeCu and NiCu from electronic structure. In
comparison with the occupation of Ni, the occupation of Fe in the
alloy is changed significantly from that of pure Fe(see DOS around 2
eV in Fig. 3). Therefore, the energy of Fe-Cu bond is much larger
than that of Ni-Cu bond, and the mixing enthalpies of CuFe are much
larger than those of CuNi. Furthermore, bond energies in alloys vary
with compositions can be also reflected by the DOS. The DOS peaks of
Ni vary with composition, while that of Fe in alloys with different
compositions are the same. These findings indicate that bond energy
$\varepsilon$ of Ni-Cu varies with composition. It can be known from
equation (2), the variation of bond energy induce
composition-dependent interaction parameter $\Omega$, which result
that the variation of mixing enthalpy with composition is not
parabolic. These are consistent with calculated results in Fig 2(a).
Hence, the complex variation trend of mixing enthalpies of NiCu
derive from the composition-dependent bond energy.

In addition, one can infer that the magnetic moment of Ni atom
varies with the composition and the magnetic moment of
Ni$_{0.25}$Cu$_{0.75}$ disappears. A more clear presentation of
magnetic moment of alloys with different composition is given in
Fig. 4. These phenomena can be well understand by using stoner
criterion which succeeds in explaining the magnetic property of
transitional metal alloys dictated by the filling of the
d-band\cite{feni}.  The Stoner criterion states that ferromagnetism
appears when the gain in exchange energy is larger than the loss in
kinetic energy. Therefore, there is always a competition between
ferromagnetic(FM) and paramagnetic(PM) solutions, and magnetic
properties are determined by the state which has lowest energy. The
DOS at the Fermi level is a key factor to affect the competition
between FM and PM state. A larger peak of DOS at Fermi level induces
a larger exchange energy and prefers to split into FM state. Fig. 5
illustrates the unpolarized DOS of NiCu and FeCu alloys. It is shown
in Fig. 4 and Fig. 5, the trend of magnetic moment of the system
well consistent with the variation of DOS at Fermi level.

For these alloys, the d electrons increase with the Cu
concentration, which result the Fermi level of Ni(Fe) be promoted.
The key point is, as shown in Fig. 5, since the DOS at Fermi level
of NiCu is much smaller than that of FeCu, the Fermi level of NiCu
change much rapidly with Cu concentration than that of FeCu which
induce that the peak of DOS of NiCu is much sensitive to the alloy
composition than that FeCu. Hence, the magnetic moment of Ni change
with the variation of DOS at Fermi level when the Cu concentration
is changed, and it disappears when the DOS at the Fermi level is 0.
Now, we can conclude that the strong asymmetry of mixing enthalpies
of NiCu with composition is induced by band shift because of  the
competition between FM and PM state.

It is found that the composition-dependent electronic interactions
of NiCu alloys also affect the structural property. As shown in Fig.
6, the calculated volume per atom of FeCu alloy well obeys the
vegard law, while that of NiCu deviates from what the vagard law
predicts.
  According to the vegard law\cite{a18}, the
lattice constant has a linear relationship with its composition,
i.e. $a(A_xB_{1-x})= x a(A) + (1-x) a(B)$. This implies that the
atom can be considered as rigid body whose volume is unchanged with
its chemical environment. However, the composition-dependent
electronic interaction in NiCu result the atomic volumes depend its
composition, which induces bowing effect.

 As shown in Fig. 7, the partial atomic pair correlation functions
 give more detailed information of the alloy structures.
 It can show that how large is the degree of distortion
 of equilibrium alloy structure from the perfect bcc structure.
For bcc structure, the numbers of 1st, 2nd, 3rd, 4th, 5th neighbors
are 8, 6, 12, 24, 8. As is shown in Fig. 7, both NiCu and FeCu
alloys are still bcc structure and the lattice just has a minor
change, but the NiCu alloy is more close to a perfect bcc structure.
Since the nominal number of minority d-band holes of Fe is more than
that of Ni by 2 electrons, the partial charge effect of Fe from Cu
is stronger than that of Ni. The bond length can be affected by
partial charge. Hence, the distance between atomic neighbors of FeCu
more sensitive to their chemical environment than NiCu, which induce
a larger distortion of FeCu from bcc.

The strain effect is also very important for the precipitated phase
in ferrite matrix.  We have calculated the strain energy with
lattice constant from 2.80 {\AA} to 2.90 {\AA} as shown in Fig. 8,
which covers the lattice constant range of bcc Fe and Cu. It is
shown that NiCu alloys have smaller strain energy at lattice of bcc
Fe, while FeCu alloys have smaller strain energy at lattice of bcc
Cu. At a lattice constant of Fe matrix with 2.83 {\AA}, the strain
energies are 6.52, 0.78, 0.51 meV/atom for Ni$_x$Cu$_{1-x}$(x=0.25,
0.5, 0.75) alloys, and  20.71, 13.51, 7.13 meV/atom for
Fe$_x$Cu$_{1-x}$(x=0.25, 0.5, 0.75). The strain energy of bcc Cu in
ferrite matrix is 21.6 meV/atom. These indicate that Ni and Fe can
lower the bcc Cu strain energy in ferrite matrix, while the Ni is
better that Fe in role of reducing strain energy.

From the energetic and structural properties of alloys, the
precipitated phase can be well understood. For HSLA steels, it was
observed that NiCu alloys at the matrix/precipitates heterophase
interfaces area\cite{a11}. The calculated results show that the
mixing enthalpies of FeCu are positive, which indicate that the FeCu
alloys are unstable and Cu atoms prefer to segregate to form
precipitates. The mixing enthalpies of NiCu alloys are negative when
Ni concentrations below 30 at. \%, indicating that NiCu alloys with
lower Ni concentration are energetic favorable. The element of Ni
appears in the matrix/precipitates heterophase interfaces area,
which is mainly induced the factors relate to strain: i) The volume
per atom of NiCu is similar to that of Fe matrix rather than pure
Cu, even rather than that of FeCu alloy, and the stain energy can be
much lower by Ni. ii) The structure of NiCu is more close to the bcc
structure, which induce little mismatch of lattice. These indicate
that NiCu alloys in the interface of matrix/precipitate can
contribute lower interface energy from reducing shear stress.
However, the experimental peak concentration of Ni is about 3 at. \%
in the Cu-rich precipitates which is lower than the Ni concentration
of most favorable NiCu alloy with concentration of 20 at. \%. This
indicates that the distribution of elements distribution may be
affected by the size effects and vibrational entropy in addition. In
RPV steels, the Ni in Cu-precipitates be rejected from the core
after thermal-aging or neutron-irradiated\cite{new,add5}, and the
experimental peak concentration of Ni is among 15-20 at. \%, which
is well consistent the composition of energetic favorable alloys
more closed. These indicate that the element distribution of bcc
precipitated phase at very initial of segregation stage can be well
understood from the energetic and structural properties.

\section{\label{sec:level1}  Conclusion}

In summary, we propose three 32-atom SQS supercells to mimic the
pair and multisite correlation functions of random CuFe and NiCu bcc
substitutional alloys which occur in Fe-Cu-Ni alloy steels as
precipitated second phase. Those SQS's are used to calculate the
mixing enthalpies, density of state, and lattice parameters of the
metastable random alloys. The results show that quasi-chemical
approach and vegard rule can satisfactorily predict the mixing
enthalpies and structure parameters of FeCu alloys but fail to
accurately yield that of NiCu. As can be obtained from the analysis
of electronic structure, the magnetism induced bond energy variation
with composition is the reason that quasi-chemical approach and
vegard rule fail to predict the properties of NiCu alloys.
Furthermore, the calculated results can well explain the previous
experimental observation of local composition of coherent
Copper-rich precipitates containing Nickel and confirm that
segregation of Ni is drove by thermaldynamic and chemical factor.
These suggest that the properties of random alloys have important
implications to better understand the multi-component precipitates.

In this work, we present the intrinsic bulk properties of metastable
bcc FeCu and NiCu random alloys to understand the structure of
Cu-rich precipitates. Since Cu-rich precipitates have important
roles on properties of alloy steels, understanding the formation
mechanism is highly desirable for altering the properties of steels
by controlling Cu-rich precipitates. Further investigation would
simulate realistic interface to calculate the interfacial energy of
precipitates/matrix and the diffusion properties of Cu and Fe atom
affected by Ni shell, which will clarify the role of Nickel on the
evolution of the Cu-rich precipitates.

The authors thank G. Xu and Prof. B. X. Zhou for critical
discussions. Y. P. Xie also thanks J. H. Yang and Prof. X. G. Gong
for insightful discussion. This work is financially supported by
National Science Foundation of China (Grant No. 50931003, 51001067),
Shanghai Committee of Science and Technology (Grant No. 09520500100)
, ¡°Shu Guang¡± project (Grant No. 09SG36) supported by Shanghai
Municipal Education Commission and Shanghai Education Development
Foundation, and Shanghai Leading Academic Discipline Project
(S30107). The computations were performed at Ziqiang Supercomputer
Center of Shanghai University and Shanghai Supercomputer Center.

%% The Appendices part is started with the command \appendix;
%% appendix sections are then done as normal sections
%% \appendix

%% \section{}
%% \label{}

%% References
%%
%% Following citation commands can be used in the body text:
%% Usage of \cite is as follows:
%%   \cite{key}         ==>>  [#]
%%   \cite[chap. 2]{key} ==>> [#, chap. 2]
%%

%% References with bibTeX database:

%\bibliographystyle{elsarticle-num}

%\bibliography{<your-bib-database>}

%% Authors are advised to submit their bibtex database files. They are
%% requested to list a bibtex style file in the manuscript if they do
%% not want to use elsarticle-num.bst.

%% References without bibTeX database:

\clearpage

\begin{table}[floatfix]
\caption{ \label{tab:table3} Atomic coordinates and occupation of
the 32 atoms bcc SQS. }
%\begin{ruledtabular}
\begin{center}
\begin{tabular}{c@{}}
\hline\hline
 Atomic coordinates   \\[0.05ex]
\hline
AB  \\[0.5ex]
A: ~0.0 -4.0 ~0.0; ~0.0 -2.0 ~0.0; -0.5 -0.5 -0.5; -1.0 -4.0 -1.0  \\[0.5ex]
A: -0.5 -1.5 -0.5; ~0.0 -1.0 ~0.0; -1.0 -3.0 -1.0; ~0.5 -4.5 ~1.5     \\[0.5ex]
A: ~0.5 -5.5 ~1.5; -1.5 -2.5 -0.5; -1.5 -3.5 -0.5; -0.5 -2.5 ~0.5  \\[0.5ex]
A: -1.0 -6.0 ~0.0; -0.5 -3.5 ~0.5; ~0.0 -3.0 ~1.0; -1.0 -5.0 ~0.0 \\[0.5ex]
B: -0.5 -4.5 ~0.5; -0.5 -5.5 ~0.5; ~0.0 -5.0 ~1.0; ~0.5 -2.5 ~0.5      \\[0.5ex]
B: ~0.5 -3.5 ~0.5; -1.5 -0.5 -1.5; -1.5 -1.5 -1.5; -1.0 -1.0 -1.0      \\[0.5ex]
B: -1.0 -2.0 -1.0; -0.5 -2.5 -0.5; -0.5 -3.5 -0.5; ~0.0 -3.0 ~0.0   \\[0.5ex]
B: ~0.0 -6.0 ~1.0; ~0.0 -4.0 ~1.0; -1.0 -3.0 ~0.0; -1.0 -4.0 ~0.0    \\[0.5ex]
\hline
AB$_3$   \\[0.5ex]
A: ~0.0 -2.0 ~0.0; ~0.5 -2.5 ~0.5; ~0.5 -3.5 ~0.5;  ~0.0 -1.0 ~0.0  \\[0.5ex]
A: -0.5 -3.5 -0.5; ~0.0 -3.0 ~0.0; ~0.0 -4.0 ~1.0; -1.5 -2.5 -0.5     \\[0.5ex]
B: ~0.0 -4.0 ~0.0; -1.5 -0.5 -1.5; -1.5 -1.5 -1.5; -1.0 -1.0 -1.0     \\[0.5ex]
B: -0.5 -0.5 -0.5; -1.0 -4.0 -1.0; -1.0 -2.0 -1.0; -0.5 -1.5 -0.5  \\[0.5ex]
B: -1.0 -3.0 -1.0; -0.5 -2.5 -0.5; ~0.0 -6.0 ~1.0; ~0.5 -4.5 ~1.5      \\[0.5ex]
B: ~0.5 -5.5 ~1.5; -1.5 -3.5 -0.5; -1.0 -3.0 ~0.0; -0.5 -2.5 ~0.5      \\[0.5ex]
B: -1.0 -6.0 ~0.0; -1.0 -4.0 ~0.0; -0.5 -3.5 ~0.5; ~0.0 -3.0 ~1.0     \\[0.5ex]
B: -1.0 -5.0 ~0.0; -0.5 -4.5 ~0.5; -0.5 -5.5 ~0.5; ~0.0 -5.0 ~1.0     \\[0.5ex]
\hline\hline
\end {tabular}
\end{center}
%\end{ruledtabular}
%\\

\end {table}

\begin{table}[floatfix]
\caption{Comparison of correlation function $\overline{\Pi}_{k,m}$
of 32 atoms bcc SQS and ideal value of random alloy.}
%\begin{ruledtabular}
\begin{center}
\begin{tabular}{c@{~~~~~~~~~}c@{~~~~~~}c@{~~~~~~}c@{~~~~~~}c}
\hline\hline
 &AB &  & AB$_3$  &   \\[0.05ex]
\hline
 &random  & SQS  &  random & SQS  \\
$\overline{\Pi}_{2,1}$ &0  & 0       &  0.25 & 0.25  \\
$\overline{\Pi}_{2,2}$ &0  & 0       &  0.25 & 0.25  \\
$\overline{\Pi}_{2,3}$ &0  & 0       &  0.25 & 0.25  \\
$\overline{\Pi}_{2,4}$ &0  & 0       &  0.25 & 0.25  \\
$\overline{\Pi}_{2,5}$ &0  & 0.17    &  0.25 & 0.04  \\
$\overline{\Pi}_{2,6}$ &0  & 0       &  0.25 & 0.25  \\
$\overline{\Pi}_{3,2}$ &0  & 0.03    &  0.13 & 0.13  \\
$\overline{\Pi}_{4,2}$ &0  & 0       &  0.06 & 0.08  \\
\hline\hline
\end {tabular}
\end{center}
%\end{ruledtabular}
\end {table}

\begin{table}[floatfix]
\caption{Total energies of bcc primitive cell of Fe, Ni and Cu as a
function of the number of k points and the cutoff energy.}
%\begin{ruledtabular}
\begin{center}
\begin{tabular}{c@{~~~~~~~~~}c@{~~~~~~}c@{~~~~~~}c@{~~~~~~}c}
\hline\hline
Monkhorst-Pack mesh  &   energy cutoff(eV) &  bcc Fe & bcc Cu & bcc Ni\\
  \hline
 $3\times3\times3 $       &280& -8.15 & -4.13 & -5.27 \\
 $5\times5\times5 $       &280& -8.12 & -3.71 & -5.33 \\
 $7\times7\times7 $       &280& -8.17 & -3.70 & -5.39\\
 $9\times9\times9 $       &280& -8.17 & -3.70 &-5.37 \\
$11\times11\times11$      &280& -8.17 & -3.70 & -5.37  \\
 $13\times13\times13$     &280& -8.17 & -3.70 & -5.37 \\
 $11\times11\times11$     &180& -6.99 & -1.97 & -4.05\\
  $11\times11\times11$    &260& -8.15 & -3.68 &-5.36 \\
 $11\times11\times11$     &300& -8.17 & -3.70 & -5.37\\
\hline\hline
\end {tabular}
\end{center}
%\end{ruledtabular}
\end {table}

\clearpage

\begin{figure}
\begin{center}
\includegraphics[scale=0.26,angle=0]{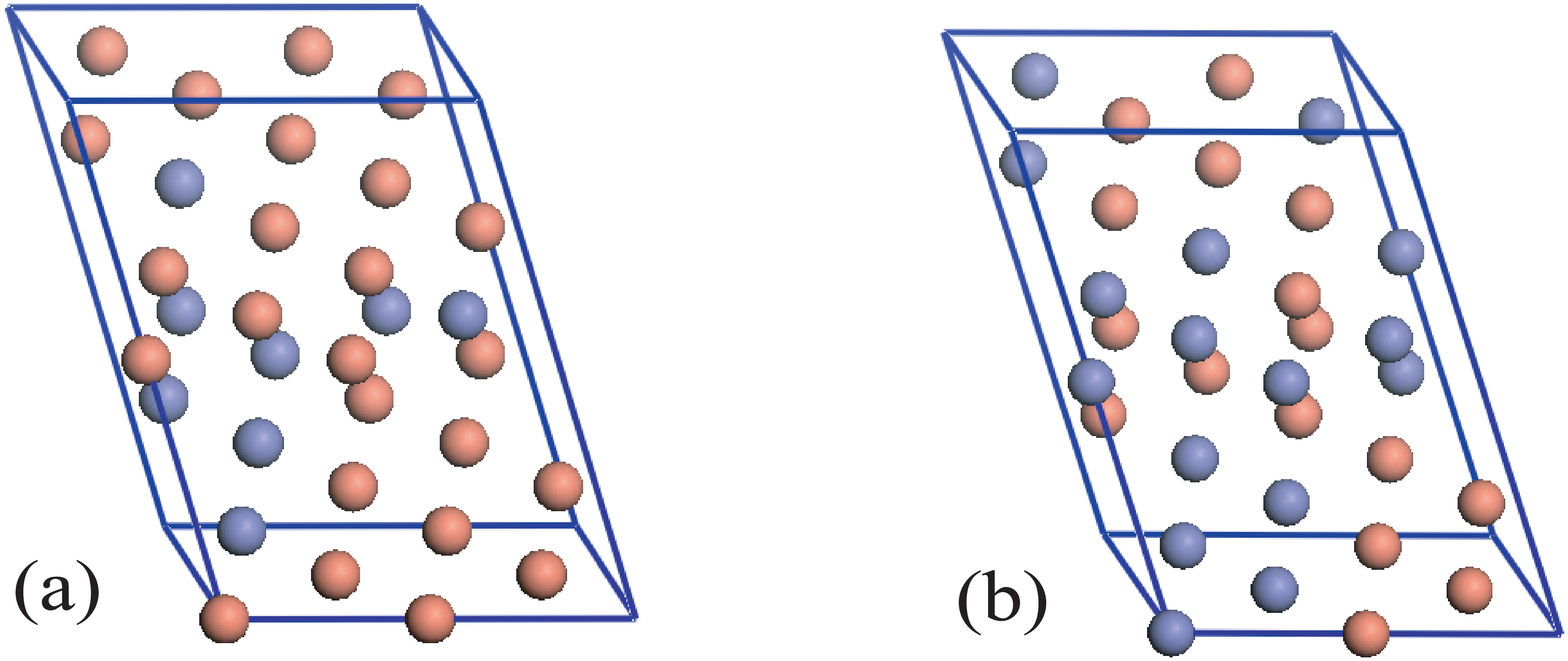}
\caption{ Crystal structures of the SQS-32 supercell in their
unrelaxed forms. Blue and red spheres represent A and B atoms,
respectively. (a) AB (b) AB$_3$ } \label{fig1}
\end{center}
\end{figure}

\begin{figure}
\begin{center}
\includegraphics[scale=0.55,angle=0]{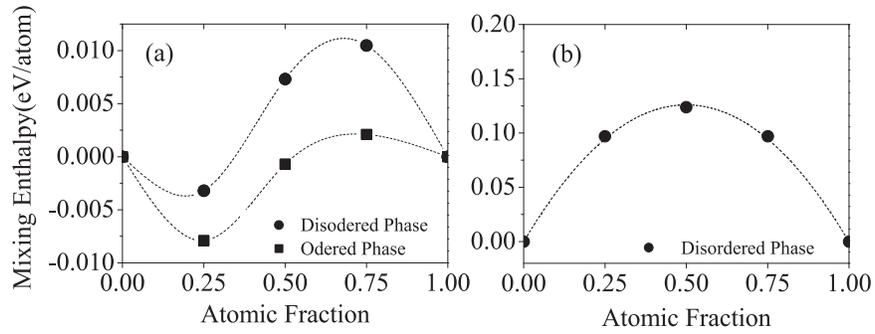}
\caption{ Mixing enthalpy of alloys: (a) disordered and ordered
Ni$_x$Cu$_{1-x}$, (b) disordered Fe$_x$Cu$_{1-x}$.} \label{fig1}
\end{center}
\end{figure}

\begin{figure}
\begin{center}
\includegraphics[scale=0.50,angle=0]{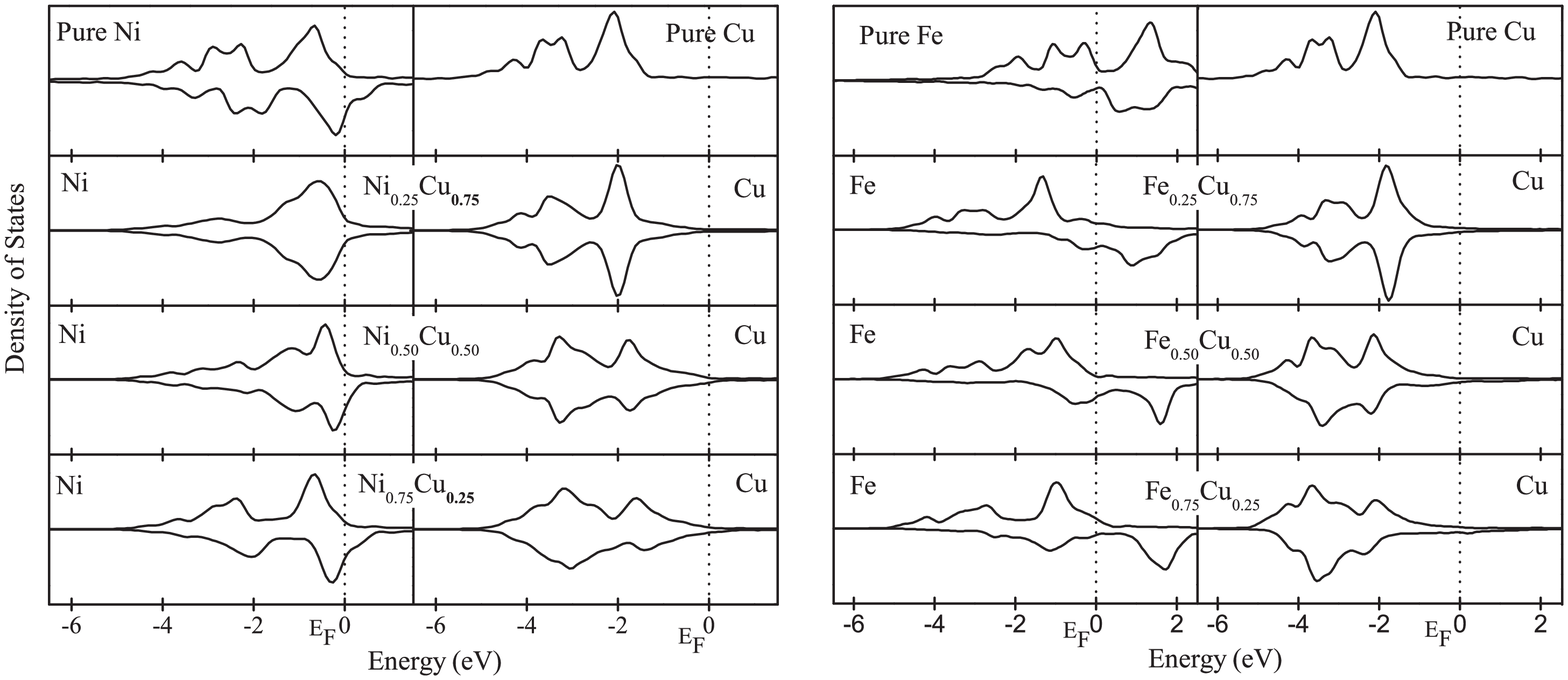}
\caption{ Comparison of the d-band between pure metals and alloys.
The d-band of Ni, Fe, Cu in pure Ni, Fe, Cu, NiCu alloy, FeCu alloy
are plot in each panel.
 } \label{fig1}
\end{center}
\end{figure}

\begin{figure}
\begin{center}
\includegraphics[scale=0.30,angle=0]{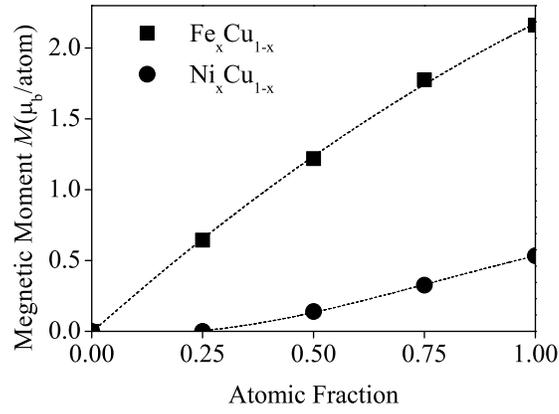}
\caption{ Magnetic moments per atom in NiCu and FeCu alloys.
   } \label{fig1}
\end{center}
\end{figure}

\begin{figure}
\begin{center}
\includegraphics[scale=0.5,angle=0]{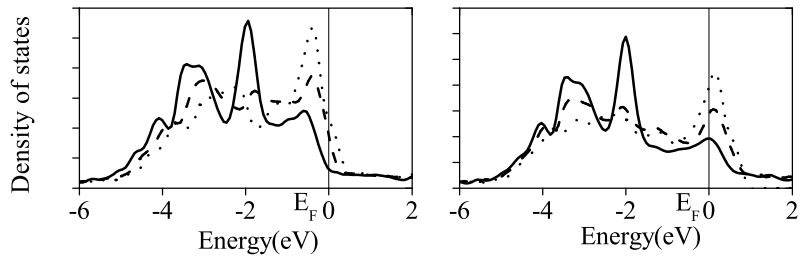}
\caption{ The unpolarized calculated density of state(DOS) of NiCu
and FeCu alloys. The solid, dash, dot line denote for
Ni(Fe)$_x$Cu$_{1-x}$, $x = 0.25, 0.50, 0.75$, respectively.
   } \label{fig1}
\end{center}
\end{figure}

\begin{figure}
\begin{center}
\includegraphics[scale=0.80,angle=0]{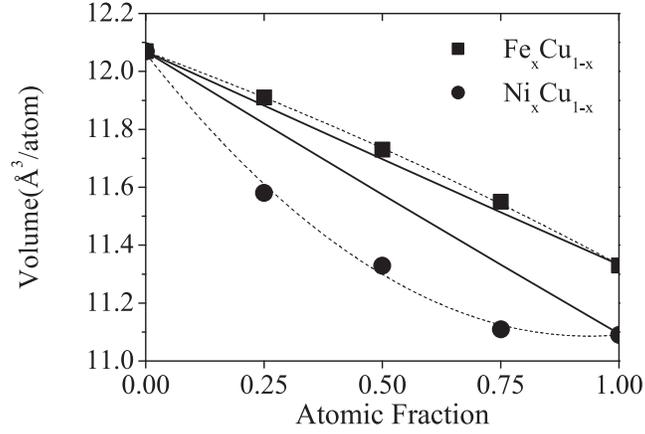}
\caption{ Atomic volume versus composition in bcc Ni$_x$Cu$_{1-x}$
and Fe$_x$Cu$_{1-x}$ alloys. The circles and squares denote the
calculated results of SQS method. The solid lines denote the results
predicted by vegard rule.
 } \label{fig1}
\end{center}
\end{figure}

\begin{figure}
\begin{center}
\includegraphics[scale=1.45,angle=0]{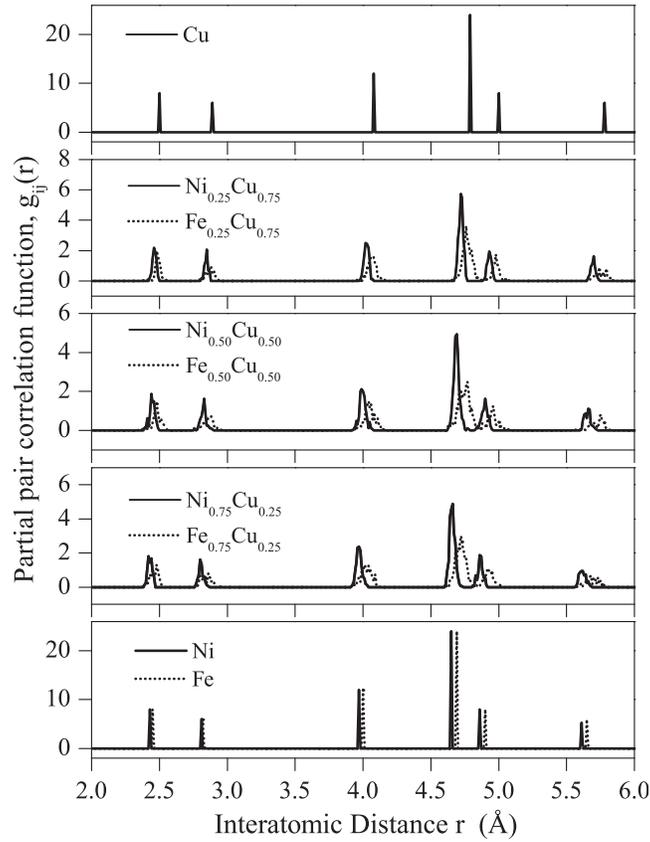}
\caption{ Partial atomic pair correlation function $g_{ij}(r)$ in
NiCu and FeCu alloys.
   } \label{fig1}
\end{center}
\end{figure}

\begin{figure}
\begin{center}
\includegraphics[scale=0.45,angle=0]{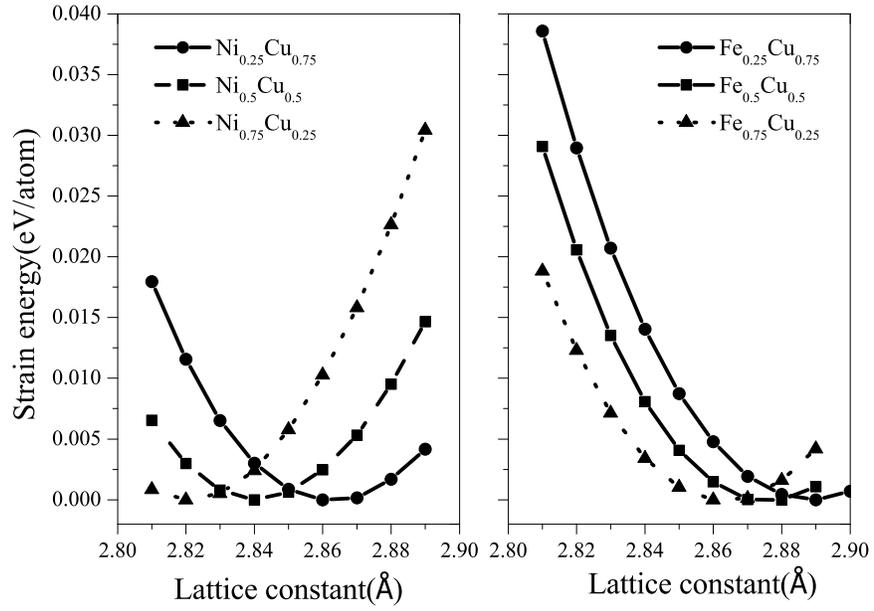}
\caption{ The strain energy of NiCu and FeCu as a function of
lattice constant.
   } \label{fig1}
\end{center}
\end{figure}

\clearpage

%\log

\end{document}